\documentclass[a4paper]{jpconf}
\usepackage{amssymb}
\usepackage{graphicx}

\def\be{\begin{equation}}
\def\ee{\end{equation}}
\def\bea{\begin{eqnarray}}
\def\eea{\end{eqnarray}}

\begin{document}
\title{Entropy accumulation near quantum critical points: effects beyond hyperscaling}

\author{Jianda Wu$^1$, Lijun Zhu$^{2}$ and Qimiao Si$^1$}

\address{$^1$Department of Physics $\&$ Astronomy, Rice University, Houston, Texas 77005, USA\\
$^2$Theoretical Division and Center for Nonlinear Studies, Los Alamos National Laboratory, Los Alamos, New Mexico 87545, USA }

\ead{jw5@rice.edu}

\begin{abstract}
Entropy accumulation near a quantum critical point was expected based on
general scaling arguments, and has recently been explicitly observed.
We explore this issue further in two canonical models for quantum
criticality, with particular attention paid to the potential effects
beyond hyperscaling. In the case of a one-dimensional transverse field
Ising model, we derive the specific scaling form of the free energy.
It follows from this scaling form that the singular temperature dependence
at the critical field has a vanishing prefactor but the singular
field dependence at zero temperature is realized. For the spin-density-wave
model above its upper critical dimension, we show that 
the dangerously irrelevant quartic coupling comes into 
the free energy in a delicate way but in the end yields only
subleading contributions beyond hyperscaling. We conclude that 
entropy accumulation near quantum critical point is a robust property
of both models.

\end{abstract}

\section{Introduction}
\label{sec:intro}

Quantum critical points (QCPs) have been extensively studied in heavy fermion
metals and related systems \cite{Si_Steglich_10}.
They occur as a non-thermal control parameter is tuned to a second-order
phase transition at zero temperature.
For thermodynamics,  it was shown \cite{Zhu}
based on scaling considerations that
the thermal
expansion [$\alpha =  (1/V)  (\partial V/\partial T)_{p,N}
\propto \partial S/\partial p$, the variation of entropy
$S$ with pressure $p$]
is more singular than the specific heat
[$c_p=(T/N) (\partial S / \partial T)_p$],
in the general case when the tuning
parameter is linearly coupled to pressure. Correspondingly,
the Gr\"{u}neisen ratio, $\Gamma =
\alpha/c_p$, diverges at the QCP,
and 
the entropy is maximized.
The same conclusions apply to a field-tuned QCP, where the magnetic
Gr\"{u}neisen ratio is the magnetocaloric effect.
The predicted divergence of the Gr\"{u}neisen ratio is by now widely
observed in quantum critical heavy-fermion metals \cite{Kuchler,Gegenwart.10},
and the entropy enhancement has recently been explicitly 
observed in a quantum critical ruthenate \cite{Rost.09}.

\begin{figure}[h!]
\begin{center}
\includegraphics[width=6.0in, height=3.0in]{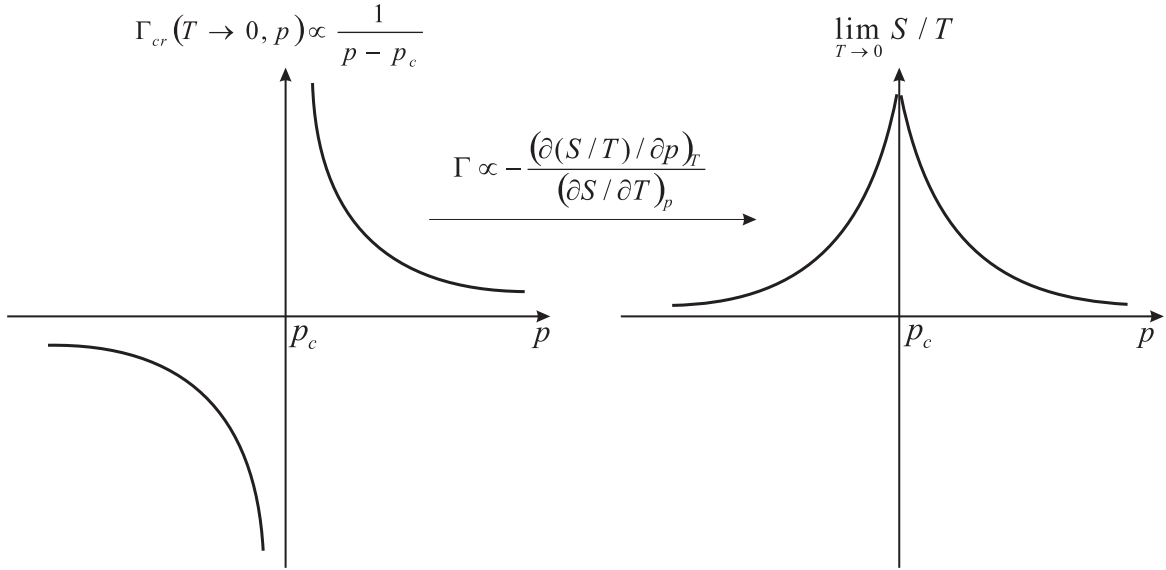}
\end{center}
\caption{Divergence of the Gr\"{u}neisen ratio and the accumulation
of entropy near QCPs.
}
\label{fig:entropy}
\end{figure}

The scaling arguments proceed as follows. Near a QCP,  the critical
part of the free energy
takes the hyperscaling form $F=F_0 T^{d/z+1} f(r/T^{1/(\nu z)})$,
where $r=p-p_c$, $d$ is the spatial dimension and $\nu$, $z$ are
respectively the correlation-length and dynamic exponents. The universal
function $f(x)$ has different asymptotic behaviors in the $x\to0$
and $x\to \pm \infty$ limits, respectively corresponding to the 
quantum critical and quantum disordered/renormalized classical
regimes.
The divergence of $\Gamma$ can be readily derived,
in universal forms as $1/r$ for $|r|\gg T$ and as $1/T^{1/{\nu z}}$
for $|r| \ll T$ \cite{Zhu}.
The $1/r$ divergence in the low-temperature limit also 
amounts to 
a sign change of $\Gamma$ 
across the QCP. As seen in Fig.~\ref{fig:entropy},
this corresponds to an entropy maximization at the QCP in the low-temperature
limit. The extension of this sign change to the finite temperature
phase transitions has also been discussed \cite{Garst}.

The hyperscaling form for the free energy is based on the existence 
of a single critical energy scale, {\it i.e.}, the gap for the
quantum critical excitations $\Delta \sim r^{\nu z}$. In this paper
we explore how the scaling form of the free energy arises in
some specific models for QCPs, paying particular
attention to the possible effects that go beyond hyperscaling.


\section{One-dimensional transverse field Ising model}

Consider first the one-dimensional transverse field Ising model
(1D TFIM), defined by the Hamiltonian \cite{Pfeuty}:
 \be H_I  =  - J\sum\limits_i {\left( {g\hat
\sigma _i^x  +  \hat \sigma _i^z \hat \sigma _{i + 1}^z } \right)},
\label{eq:ham-tfim}
 \ee
where $J>0$ is the ferromagnetic Ising coupling
between 
the neighboring spins and $gJ$ is the transverse magnetic field. With
the tuning of $g$, there exists a quantum phase transition  from
a ferromagnet to quantum paramagnet at $g_c=1$ \cite{Pfeuty,chakrabarti}.
Using a Jordan-Wigner transformation, we can represent the
Hamiltonian in the thermodynamic limit in terms of free fermions 
and the free energy can be written
as
$ F =   - {Nk_B T}\left[ \ln 2 + \frac{1}{\pi }\int_0^\pi
d\;k \ln \cosh  (\varepsilon_k  /  2k_B T ) \right]$,
with the dispersion $\varepsilon _k = 2J \left( {1 + g^2 - 2g\cos k}
\right)^{1/2}$. The normalized free energy density is
 \be f \equiv \frac{F}{{NJk_B }} =  f_0(g) - \frac{t}{\pi }\int_0^\pi {dk\ln
\left( {1  + e^{ - 2A} } \right)}\, ; A = \sqrt {
(1-g)^2/t^2
+4
(g/t^2) {\sin^2 \left( {k/2} \right)}}\;
 \label{eq:free-tfim-qc}
 \ee
where $t=k_B T/J$, and $f_0(g)$ is the ground state energy.
Our focus will be on the $T$-dependent part, 
$f_1(g,T)=f (g,T) - f_0(g)$.

Consider first the quantum critical regime, where
$\left| {g - 1} \right|/t \ll 1$. 
At temperatures small compared to $J$, 
we can introduce a momentum cutoff $k_c \approx t$, below
which $\left| {\sin \left( {k/2} \right)} \right|/t \ll 1$
and therefore $A \ll 1$. The free
energy can be approximated by 
the contributions from $k<k_c$, which
is
\bea
f_1 &\approx & - \frac{t}{\pi }\left[ {\frac{1}{2}\left( {\frac{{1 - g}}{t}} \right)^2 k_c
  + \frac{g}{{t^2 }}\frac{{k_c^3 }}{6} + k_c \ln 2} \right]  
 \mathop  \approx \limits^{k_c  \approx t}   - \frac{{t^2 }}{\pi
}\left( {\left( {\frac{g}{6} + \ln 2} \right) + \frac{1}{2}\left(
{\frac{{1 - g}}{t}} \right)^2 } \right)\;.
\label{tfim_qc}
 \eea
While Eq.~(\ref{tfim_qc}) 
contains approximations for the regular pieces,
it reveals an important point.
Compared with the general hyperscaling form, this expression
is particular in that 
the linear term $|1-g|/t$ in the expansion vanishes.
As a result, the magnetic Gr\"{u}neisen ratio
as a function of temperature at the QCP 
stays finite.

Consider next the low-temperature regions where  $\left| {1 - g} \right|/t 
\gg 1$. Here $A \gg 1$ and we can expand $\ln(1+e^{-2A})$ as $e^{-2A}$ 
and obtain
\be
f_1
\approx  - \frac{{t^2 }}{{\sqrt {2\pi g} }}\left( {\frac{{\left| {1 - g} \right|}}{t}} \right)^{1/2} e^{ - \frac{2}{t}\left| {1 - g} \right|} \;.
\label{eq:fs-tfim-qd}
 \ee
We can further replace 
$1/\sqrt{g}$ by $1/\sqrt{g_c}$ without missing any 
singularity.
The free energy has the exact
hyperscaling form with $z=1$.  We obtain the magnetic Gr\"{u}neisen ratio to be
\be
\Gamma = \frac{1}{g-1}  \;,
\label{eq:gruneisen-tfim-qd}
 \ee
which is indeed divergent at the QCP, $g_c=1$. 
The accumulation of entropy immediately follows (Fig.~\ref{fig:entropy}).

\section{Spin-density-wave quantum critical point}
\label{sec:sdw}

Quantum phase transitions in itinerant magnets are traditionally described 
in terms of a $T=0$ spin-density-wave (SDW)
transition and formulated as a quantum Ginzburg-Landau theory with dynamic
exponent $z>1$~\cite{hertz}.
For concreteness, we take the 3D antiferromagnetic case
as an example, {\it i.e.}, with $d=3$ and $z=2$.
The effective dimension of the Ginzburg-Landau theory is $d+z$,
which is above the upper critical dimension $4$. 
Correspondingly, the quartic coupling
$u$, appearing in the action as $u \int \phi^4$, is irrelevant in the
renormalization-group (RG) sense.
Under the RG transformation, $u$ renormalizes to zero as the fixed 
point is reached,
leaving only the Gaussian (quadratic) part.
However, $u$ is dangerously irrelevant~\cite{millis,Moriya}.
For instance, in the quantum critical regime,
the correlation length is determine 
as $\xi^{-2} \sim r(T) = r + c u T^{3/2}$ 
(where $c$ is a constant,
$4(n+2)(1/3 - \ln 2 \sqrt 2 /(3\pi^3))$, with 
$n$ being the number of components of the order parameter).
One can consider $u$ as introducing a new energy scale,
and the scaling functions for generic physical quantities
are expressed in terms of two variables $f(r/T, u T^{3/2}/r)$.

Within the RG approach \cite{millis},
it is natural to focus on the free energy 
associated with the Gaussian term, $F_{G}$.
The expression for $F_{G}$ in the RG approach can be shown
to be equal to a Gaussian free energy 
with $r$ replaced by renormalized $r(T)$.
We will focus on the quantum critical regime, and write 
$F_{G} = F_{G}^{(1)}+
F_{G}^{(2)}$,
where
 \bea
F_{G}^{(1)} =
 - nV\int_0^\Lambda  {\frac{{d^3 q}}{{(2\pi )^3 }}\int_0^{\Gamma _q }
{\frac{d \varepsilon }{{2\pi }} \left ( \coth \frac{\varepsilon }{{2T}} -1 \right )
\tan ^{ - 1} \frac{{\varepsilon /\Gamma _q }}{{r(T) + (q/\Lambda )^2 }}} }
\;,
 \label{FG1}
 \eea
and
 \bea
 F_{G}^{(2)} =
- nV\int_0^\Lambda  {\frac{{d^3 q}}{{(2\pi )^3 }}\int_0^{\Gamma _q }
{\frac{d \varepsilon }{{2\pi }}
\tan ^{ - 1} \frac{{\varepsilon /\Gamma _q }}{{r(T) + (q/\Lambda )^2 }}} }
\;.
 \label{FG2}
 \eea
Here, $\Gamma_{q} = \Gamma_0 q ^{z-2}$,
while $\Lambda$ and $\Gamma_0$ are respectively 
the ultraviolet momentum and energy cutoffs.
The leading temperature-dependent contribution from
$F_{G}^{(2)}$ turns out to be
$\frac{{nV\Lambda ^3 \Gamma _0 }}{{4\pi ^3 }}\left( 
{\frac{{\sqrt 2 \pi }}{4} + \ln 2 - \frac{{\sqrt 2 }}
{2}\ln (\sqrt2 +1)} \right)
c (u/\Gamma_0^{3/2})T^{3/2}$, which is more singular than that from
$F_{G}^{(1)}$.

This Gaussian form, however, is incomplete. 
The quartic coupling also introduces an explicitly linear
in $u$ term ($t \equiv T/\Gamma_0$), which
takes the form
 \bea
F_u  =  - 2n(n + 2)N\Gamma _0 u\int_0^\infty  
{e^{ - 6x} \left( {\int_0^1 {\frac{{d^3 q}}{{(2\pi )^3 }}}}
\int_0^1 {\frac{{d\varepsilon }}{\pi }}
\coth \frac{\varepsilon }{{2te^{2x} }}
\frac{\varepsilon }{{\varepsilon ^2  + (re^{2x}  + q^2 )^2 }} \right)~f_u}
~d x
\;,
\nonumber
\\
f_u \equiv \frac{1}{2\pi ^2 }
\int_0^1 \frac{d\varepsilon }{\pi }
\coth \frac{\varepsilon }{2te^{2x} }
\frac{\varepsilon }{\varepsilon ^2  + (re^{2x}  + 1)^2 }
+ \frac{2}{\pi }\int_{}^{} \frac{d^3 q}{(2\pi )^3 }
\coth \frac{1}{2te^{2x} }\frac{1}{{1 + (re^{2x}  + q^2 )^2 }}
\;.
 \label{F_u}
 \eea
We find that, to the linear order in $u$, the leading temperature-dependent 
contribution from $F_u$ exactly cancels that of $F_{G}^{(2)}$, leaving the
singular contribution to the total free energy to be
 \bea
F_{tot} \approx F_{G}^{(1)}
\;,
 \label{Ftot}
 \eea
which is given by 
 \bea
F_{tot} 
\sim
- 2\frac{{T^2 }}{{\Gamma _0^2 }}
   + \frac{{2\pi T^{5/2} }}{{\Gamma _0^{5/2} }}\left( {1 + \frac{9}{2}\frac{r + cu T^{3/2}/\Gamma_0^{3/2}}{{T/\Gamma _0 }}} \right)
\;.
 \label{Ftot_final}
 \eea

We can trace the singular terms included in Eq.~(\ref{FG1})
to the contributions of the low-energy degrees
of freedom, and those included in Eq.~(\ref{FG2})
to a unphysical origin 
from the degress of freedom at high energies and 
short distances near the cutoff scales.
Eqs.~(\ref{FG1},\ref{Ftot}) are therefore expected to be the 
regularization prescription that is valid to all orders in $u$.
The explicit result given in Eq.~(\ref{Ftot_final}) is consistent 
with those often quoted in the literature.
The $T^2$ term
represents 
the background
Fermi liquid
contribution. The critical part 
of the free energy
is consistent with the scaling form
$T^{5/2}(1 + r/T)$. 
The contribution of the renormalized 
quartic coupling $uT^{3/2}$ to the
free energy is subleading; the dangerously irrelevant parameter 
does not modify the leading singular part of the
free energy from its hyperscaling form,
thereby preserving the 
divergence of the 
Gr\"{u}neisen ratio and the accompanied entropy accumulation.

\section{Summary}
\label{sec:summary}

We have considered the thermodynamic properties of two models
of quantum criticality. We have analyzed the potential effects beyond the
most general hyperscaling forms in both models, and established
that entropy accumulation is a robust property of both models.

\ack
We acknowledge the support of
NSF Grant No. DMR-1006985, the Robert A. Welch Foundation
Grant No. C-1411 (JW and QS), and the U.S. DOE through the LDRD program
at LANL (LZ).

\end{document}